\documentclass[%
aip,
 amsmath,amssymb,
 reprint,%
]{revtex4-1}

\usepackage{graphicx}
\usepackage{dcolumn}
\usepackage{bm}

\usepackage[utf8]{inputenc}
\usepackage[T1]{fontenc}
\usepackage{mathptmx}
\usepackage{etoolbox}
\usepackage{color,soul}

\makeatletter
\def\@email#1#2{%
 \endgroup
 \patchcmd{\titleblock@produce}
  {\frontmatter@RRAPformat}
  {\frontmatter@RRAPformat{\produce@RRAP{*#1\href{mailto:#2}{#2}}}\frontmatter@RRAPformat}
  {}{}
}%
\makeatother
\begin{document}

\preprint{AIP/123-QED}


\title[Manuscript]{Chemically reactive and aging macromolecular mixtures II: Phase separation and coarsening}
\author{Ruoyao Zhang}
\affiliation{ 
Department of Mechanical and Aerospace Engineering, Princeton University, Princeton, New Jersey 08544, United States
}
\author{Sheng Mao}%
\affiliation{ 
Department of Mechanics and Engineering Science, College of Engineering, Peking University, Beijing 100871, China
}%

\author{Mikko P. Haataja}
\affiliation{
Department of Mechanical and Aerospace Engineering, Princeton University, Princeton, New Jersey 08544, United States
}%
\affiliation{
Princeton Materials Institute, Princeton University, Princeton, New Jersey 08544, United States}%
\affiliation{
Omenn-Darling Bioengineering Institute, Princeton University, Princeton, New Jersey 08544, United States}
\email{maosheng@pku.edu.cn, mhaataja@princeton.edu}

\date{\today}

\begin{abstract}
In a companion paper, we put forth a thermodynamic model for complex formation via a chemical reaction involving multiple macromolecular species, which may subsequently undergo liquid-liquid phase separation and a further transition into a gel-like state.  
In the present work, we formulate a thermodynamically consistent kinetic framework to study the interplay between phase separation, chemical reaction and aging in spatially inhomogeneous macromolecular mixtures. 
A numerical algorithm is also proposed to simulate domain growth from collisions of liquid and gel domains via passive Brownian motion in both two and three spatial dimensions. 
Our results show that the coarsening behavior is significantly influenced by the degree of gelation and Brownian motion. The presence of a gel phase inside condensates strongly limits the diffusive transport processes, and Brownian motion coalescence controls the coarsening process in systems with high area/volume fractions of gel-like condensates, leading to formation of interconnected domains with atypical domain growth rates controlled by size-dependent translational and rotational diffusivities.
\end{abstract}

\maketitle

\section{\label{sec:level1} Introduction}

\begin{figure*}[!htb]
\includegraphics[width=0.85\linewidth]{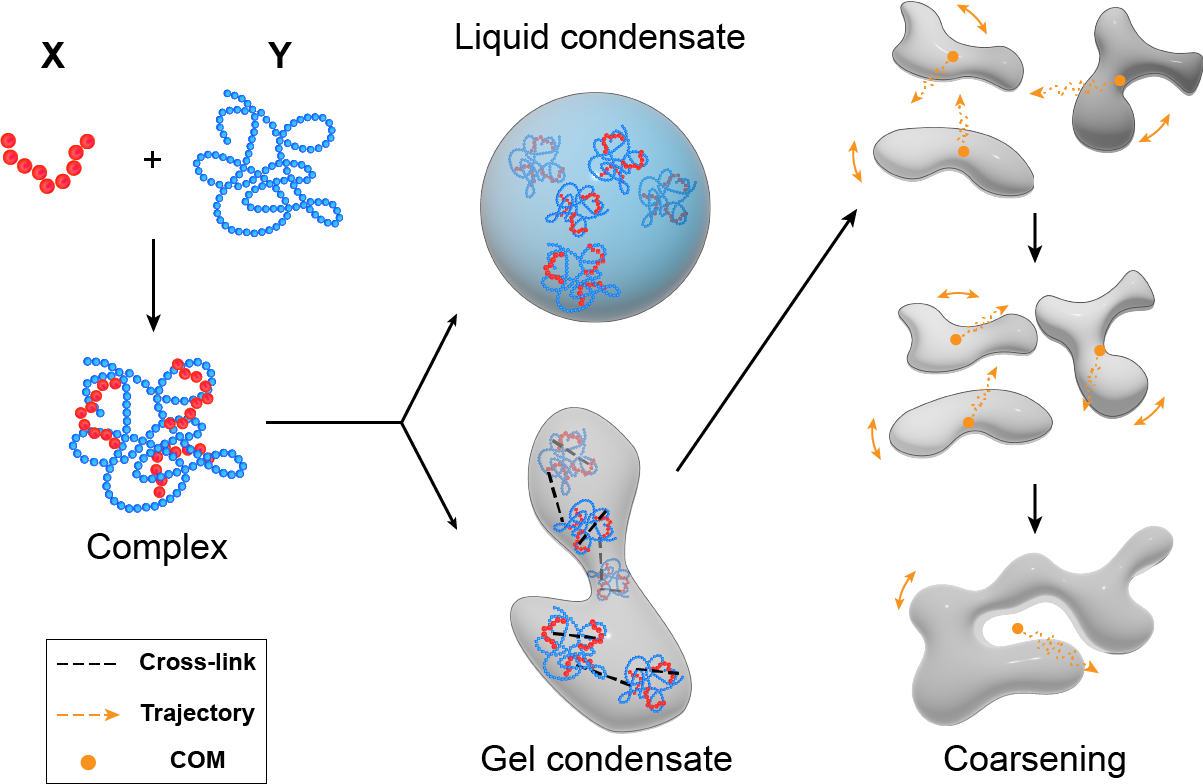}
\caption{\label{fig:schematic} Schematic of the Brownian motion-driven coarsening mechanism considered in this work.  Two macromolecular species $X$ and $Y$ react to form a complex, $Z$, which subsequently phase separates from the buffer to form condensates. Liquid-like condensates coarsen via combination of diffusion-limited coarsening (DLC) and Brownian-motion-induced coalescence (BMC) while preserving spherical morphology. Gel condensates appear anisotropic due to the presence of physical cross-links, while their Brownian motion leads to the formation of irregular, possibly interconnected structures.}
\end{figure*}

While multi-component liquid-liquid phase separation (LLPS) in various physical systems has been broadly studied both theoretically and experimentally \cite{Brangwynne_2009, Hyman_2014_review, Berry_2018, Deviri2021}, possible chemical reactions between the components introduce another layer of complexity, changing the underlying microscopic picture. The product of the chemical reactions between different macromolecular species is often referred to as a complex, which in turn may form mesoscopic aggregates or biomolecular condensates via LLPS. For instance, oppositely charged polymers react with one another to form ``complex coarcervates'', exhibiting physical properties different from the individual polymers \cite{Lu2020, Sing2020}. Similar phenomena have also been observed in other systems, such as colloid-polymer blends, proteins and nucleic acids \cite{Poon_2002, Alberti_review_2017, Perry2019}. Theoretical models are able to address many questions about reaction-induced LLPS, and predict phase behaviors in simple systems \cite{Corrales_Wheeler, Talanquer, RADHAKRISHNAN_PNAS, Weber2019, Kirschbaum_2021}. 

The condensates emerging via LLPS do not always behave liquid-like, however, as physical cross-links can form  between individual macromolecules~\cite{TanakaT, TanakaF, Harmon2017}. This process, often interpreted as aging or gelation, restricts the mobility of condensates, resulting in a slower kinetics~\cite{SHIN_science, Jawerth_Science}. Although gel phases may contribute to the biological functions during protein assembly~\cite{Putnam2019}, further transitions from the gel phase into a more solid-like phase (e.g., fibrillar aggregates) are often associated with various neurodegenerative diseases~\cite{PATEL20151066, PESKETT2018588, StGeorgeHyslop2018, Michaels_review_2023}. Therefore, to tackle the problem of complex formation and unravel its relation to LLPS, a thermodynamic model for multi-component chemically reactive macromolecular mixtures was formulated in a companion paper \cite{Zhang_part1}. Ternary phase diagrams were constructed to predict phase behavior of quaternary mixtures that may undergo chemical reaction together with LLPS as well as their propensity to further transition into a gel-like state. 

Importantly, chemical reactions and spatially inhomogeneous compositions often drive a multi-component system out of macroscopic equilibrium, which cannot be addressed only by the thermodynamic model without explicitly considering the kinetics. A case in point is the Brownian motion of small particles or aggregates suspended in a fluid~\cite{Karatzas_1998}. It has been shown that in an aqueous solution or a crowded cellular environment, Brownian motion will lead to displacements of micrometer-sized structures~\cite{Berg_1984, DiRienzo_BM_2014, Wang_BM_2019}. Colloidal systems, in which individual particles exhibit Brownian motion, can form clusters and gels~\cite{Sedgwick_gelBM_2004, Zaccarelli_ColloidReview_2007}. It has also been shown that diffusion-limited coarsening (DLC) and Brownian-motion-induced coalescence (BMC) contribute significantly to the coarsening kinetics of condensates~\cite{Forteln_BMC_1995, Marsh_DLC_1996, Berry_2018}. Berry et al.~studied the coarsening of liquid-like biomolecular condensates (nucleoli) in three dimensions (3D) and found the dominance of BMC at late times~\cite{Berry_2015}. 
Wilken et al.~in turn performed two-dimensional (2D) simulations to study condensates formed by DNA nanostars and investigated their near-equilibrium Brownian motion~\cite{Wilken_2023}.

To the best of our knowledge, however, the existing mesoscopic models of multi-component LLPS or reaction-induced phase separation do not capture the kinetic aspect of the aging of coacervates/condensates and their Brownian motion.
To bridge this knowledge gap, herein we develop a thermodynamically consistent approach to study the spatio-temporal evolution of multi-component, chemically reactive macromolecular mixtures, and examine the interplay between LLPS, chemical reactions, aging (gelation), and Brownian motion operating concurrently. As detailed in a companion paper~\cite{Zhang_part1}, our thermodynamic model predicts phase behavior of a macromolecular mixture and its propensity to undergo aging via gelation, which can be affected by molecular sizes, stoichiometric coefficients, equilibrium constants and interaction parameters.
In this paper, with an emphasis on the kinetic aspect of such mixtures, our results imply that the coarsening behavior is significantly influenced by the degree of gelation and Brownian motion. The presence of a gel phase inside condensates strongly limits diffusive transport processes, and BMC controls the coarsening process in systems with high area/volume fractions of gel condensates, leading to the formation of interconnected gel domains with atypical growth rates controlled by size-dependent translational and rotational diffusivities.

\section{Kinetic model}

\subsection{Thermodynamic free energy}

The thermodynamic model of a homogeneous macromolecular mixture that can undergo LLPS and further transition into gel-like state was presented in our companion paper~\cite{Zhang_part1}. Here, we adopt the formulation of a binary reversible chemical reaction between two molecular species $X$ and $Y$ in forming a complex $Z$ inside a buffer solution $B$, i.e., 
\begin{equation}
    nX + mY \leftrightarrow X_{\mathrm{n}} Y_{\mathrm{m}} \equiv Z,
\end{equation}
 where $n$ and $m$ are the stoichiometric coefficients. The liquid free energy density is \cite{RADHAKRISHNAN1999, RADHAKRISHNAN_PNAS}
\begin{equation}
\label{eqn:fliquid}
\begin{aligned}
&f_{\mathrm{liquid}}(X,Y,Z) =\frac{X}{r_x} \ln X+ \frac{Y}{r_y} \ln Y+ \frac{Z}{r_z}\left(\ln Z + \mu ^0 _z\right) \\
&+ B\ln B +\chi _{xy}XY+\chi _{xz}XZ + \chi _{yz} YZ \\
&+ \chi _{xb}XB +\chi _{yb}YB + \chi _{zb}ZB,
\end{aligned}
\end{equation}
where $r_x$, $r_y$ and $r_z$ denote degrees of polymerization; $\chi_{ij}$'s are the interaction parameters between species $i$ and $j$; and $\mu ^0_z=-\ln K$ denotes the standard chemical potential of the complex, where the parameter $K$ corresponds to the equilibrium constant for the reaction in an ideal solution ($\chi_{ij}=0$) with equal degrees of polymerization~\cite{Zhang_part1}. It should be mentioned that if one considers only the change in the translational entropy associated with the complexes, $r_z = n r_x + m r_y$; however, to allow for possible changes in chain conformations upon complex formation, $r_x$, $r_y$ and $r_z$ can be treated as independent parameters.

Now, as far as gelation is concerned, a simple criterion based purely on the local complex volume fraction was employed in our previous work~\cite{Zhang_part1}, which enabled the identification of the gelation regions in phase diagrams by simply constructing iso-contour lines of the complex volume fraction $Z$. In the present work, the gelation criterion is re-formulated with the aid of a gelation order parameter $\phi \in [0, 1]$ for convenience. To this end, the liquid-to-gel transition within the clusters of molecular complexes is captured using the simple free energy density~\cite{Sciortino}
\begin{equation}
\label{eqn:fgel}
f_{\mathrm{gel}}(Z, \phi) = f_{\mathrm{g}}\left[-\frac{g(Z)}{2} \phi^{2}+\frac{\phi^{3}}{3}\right],
\end{equation}
where $f_g>0$ denotes a characteristic energy density scale, and the term $g\left(Z\right)$ couples gel concentration to the complex volume fraction via
\begin{equation}
\label{eqn:gcomplex}
g(Z)=\frac{pZ - Z ^ *}{1-Z ^*},
\end{equation}
where
\begin{equation}
p=\frac{\exp \left(\Delta F_c / k_{\mathrm{B}} T\right) }{1+\exp \left(\Delta F_c / k_{\mathrm{B}} T\right)}.
\end{equation}
The parameter $p$ denotes the fraction of the monomers in the polymer which are in the proper configuration to form cross-links, such that $pZ$ is the volume fraction of cross-links in the system. 
$\Delta F_c$ in turn denotes the change in free energy when forming a cross-link in the chain; $k_\mathrm{B}$ and $T$ are Boltzmann constant and temperature, respectively, while $Z^\ast$ denotes the critical cross-linking volume fraction necessary to form a gel.  The form of Eq.~\ref{eqn:fgel} ensures that when $pZ > Z^\ast$, $f_{\mathrm{gel}}$ has a local minimum at $\phi = g(Z)\in (0,1)$, while for $pZ < Z^\ast$, the local minimum of $f_{\mathrm{gel}}$ is at $\phi= 0$.
Together, the total free energy of the system is thus written as
\begin{equation}
F= \int_V d^3\vec{r} \left[ f_{\mathrm{liquid}}(X,Y,Z) + f_{\mathrm{gel}}(Z,\phi)\right],
\label{eq:bulk}
\end{equation}
where $V$ denotes the system volume. Equation (\ref{eq:bulk}) also describes the bulk free energy of all molecular species, and minimizing it provides one with the thermodynamic driving forces that govern the system's evolution.

\subsection{Spatially inhomogeneous, chemically reacting systems}

The thermodynamic model only describes the properties of a spatially uniform mixture at equilibrium. However, the spatio-temporally varying compositions and their kinetics often drive the system out of equilibrium, thus necessitating the development of a model which properly incorporates capillary (i.e., interfacial) effects, mass transport kinetics and chemical reactions. To account for capillary effects, we introduce standard squared gradient terms to the total free energy:
\begin{equation}
\begin{aligned}
F&=\int _V d^3 \vec{r}\left\{\frac{\epsilon_{\phi}^{2}}{2}|\vec{\nabla} \phi|^{2}+\frac{\epsilon_{x}^{2}}{2}|\vec{\nabla} X|^{2}+ \frac{\epsilon_{y}^{2}}{2}|\vec{\nabla} Y|^{2} \right. \\
&\left. + \frac{\epsilon_{z}^{2}}{2}|\vec{\nabla} Z|^{2} + f_{\mathrm{liquid}}(X,Y,Z) + f_{\mathrm{gel}}(Z,\phi)\right\},
\end{aligned}
\end{equation}
where $\epsilon _\phi$, $\epsilon _x$, $\epsilon _y$ and $\epsilon _z$ denote the gradient energy coefficients setting the scale of the interfacial tensions for $\phi$, $X$, $Y$ and $Z$, respectively \cite{Provatas_2011}. 

In describing the time evolution of $X$, $Y$, and $Z$, it is important to recognize that an undriven reactive system at equilibrium must not only satisfy phase equilibria (equality of chemical potentials between coexisting phases), it must also satisfy chemical equilibria~\cite{Berry_2018}. To achieve this, we construct our kinetic equations based on a thermodynamically consistent theory of undriven chemically reactive, inhomogeneous systems set forth by Bazant \cite{Bazant_2013}. Furthermore, we adopt the form of chemical potentials derived by Kirschbaum and Zwicker~\cite{Kirschbaum_2021} for species with different molecular volumes undergoing chemical reactions. Thus, our kinetic equations for conserved molecular species $X$, $Y$ and $Z$, are written as combinations of Cahn-Hilliard equations (Model B) \cite{Hohenberg_1977} plus terms emanating from chemical reactions:
\begin{equation}
\label{eqn:dxdt}
\frac{\partial X}{\partial t}=\nabla \cdot \left( M_{x} v_x\nabla \frac{\delta F}{\delta X} \right) - nv_xR, 
\end{equation}
\begin{equation}
\label{eqn:dydt}
\frac{\partial Y}{\partial t}=\nabla \cdot \left( M_{y} v_y \nabla \frac{\delta F}{\delta Y} \right) - mv_yR, 
\end{equation}
\begin{equation}
\label{eqn:dzdt}
\frac{\partial Z}{\partial t}=\nabla \cdot \left( M_{z} v_z \nabla \frac{\delta F}{\delta Z} \right) + v_zR, 
\end{equation}
where $M_i$ denotes the mobility constant for species $i$.  Volume conservation during chemical reactions is enforced by imposing $v_z = nv_x + mv_y$, which is consistent with the formulation in the thermodynamic model. Furthermore, we adopt the following form of rate equation for chemical reactions, which is consistent with both detailed balance and non-equilibrium thermodynamics~\cite{Bazant_2013}:
\begin{equation}
R =\tilde{k}_{0} \left[ \exp \left(\frac{nv_x}{k_{\mathrm{B}} T} \frac{\delta F}{\delta X}+\frac{mv_y}{k_{\mathrm{B}} T} \frac{\delta F}{\delta Y}\right) -\exp \left( \frac{v_z}{k_{\mathrm{B}} T} \frac{\delta F}{\delta Z}\right) \right],
\end{equation}
where $\tilde k_0$ denotes a characteristic reaction rate. 
It is easy to verify that as the system is driven towards phase and chemical equilibria given by $\mu_i = \delta F/\delta X_i = const.$ and $n v_x \mu_x + m v_y \mu_y = v_z \mu_z$, respectively, $R=0$ and $\partial X_i/\partial t = 0$ by construction.

Finally, the gelation process in turn can be interpreted as a non-conserved phase transformation. The time evolution of the local gel order parameter (OP) $\phi\left(\vec{r},t\right)$ is governed by the Allen-Cahn equation (Model A) \cite{Hohenberg_1977, Sciortino}:
\begin{equation}
\label{eqn:dphidt}
\frac{\partial \phi}{\partial t}=-M_{\phi} \frac{\delta F_{\phi}}{\delta \phi}, 
\end{equation}
where $M_\phi$ denotes the mobility constant of the gel phase. In addition, we assume that the rate of chemical reactions and complex mobility decrease drastically within the gel phase due to the existence of physical cross-links, i.e., $\tilde{k}_0(\phi) = k_0 \exp\left(-\phi/\phi_0\right)$ and $M_z (\phi) = M_{z,0} \exp\left(-\phi/\phi_0\right)$, where $k_0$ and $M_{z,0}$ denotes the reaction constant and the mobility of the complex $Z$ at liquid state, and $\phi_0 \ll 1$ denotes a characteristic gel OP value above which the reaction slows down significantly. That is, the formation and diffusion of the complexes are limited by the presence of the gel phase, which may force the system to remain in a non-equilibrium state over macroscopic time scales.

\section{Numerical methods}

\subsection{Phase field solver}

To numerically study the time-dependent morphologies and aggregate size distributions that emerge from the model, the fields $X$, $Y$, and $Z$ are initialized with small fluctuations around their respective initial values ($X_0$, $Y_0$ and $Z_0$) within the spinodal region according to the ternary phase diagrams constructed in paper I \cite{Zhang_part1}. The gel OP $\phi$ is also initialized with small fluctuations. We then numerically integrate Eqns.~(\ref{eqn:dxdt}-\ref{eqn:dzdt}) and (\ref{eqn:dphidt}) using a forward in time, centered in space scheme on a uniform square/cubic grid with periodic boundary conditions and edge length $L$. During each iteration, the reaction rate $R$ is evaluated using the solutions from the previous time step and then coupled to Model B equations to yield the new fields  $X$, $Y$, and $Z$. The Laplacians in 2D (3D) are evaluated using isotropic 9-point (27-point) stencils \cite{Patra_2006}. 

In the simulations, we employ the following non-dimensional parameter values: $N_x=N_y=N_z=200$, $\Delta x=\Delta y=\Delta z=1$, $\Delta t=1\times{10}^{-4}$, $M_x=M_y=M_{z,0}=1$, $k_0=1$, $\epsilon _x =\epsilon _y = \epsilon _z = 8$, $\epsilon _\phi = 0.01$, and $\phi_0=0.001$. These parameter values were chosen to produce physically reasonable outputs, as our aim is to study the general behavior of complex formation and phase transformations rather than simulate a particular macromolecular mixture.

\begin{figure*}[!ht]
\includegraphics[width=1\linewidth]{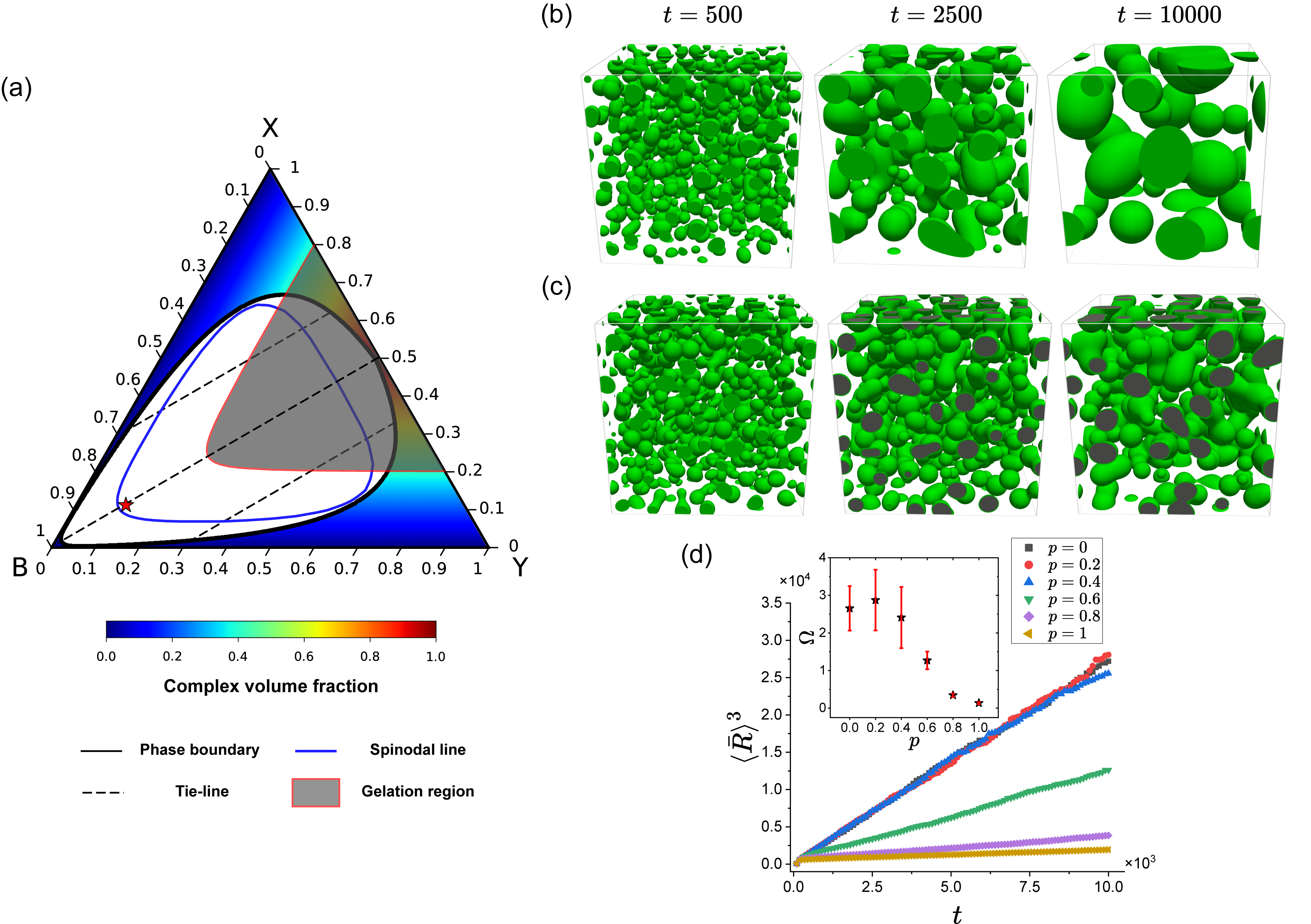}
\caption{\label{fig:3D_noBM} Coarsening behavior in 3D systems without Brownian motion. (a) Phase diagram of a representative system. Parameters employed: $n=m=1$, $v_x=v_y=1$, $r_x=r_y=r_z=1$, $\chi_{zb}=4$, $K=100$, $Z^\ast=0.4$ and $p=1$. For $p<1$, the gelation region shrinks and disappears altogether when $p<Z^\ast$.  Red star: initial condition at $(X_0 = Y_0=0.118)$. (b) Time sequence of liquid-like condensate configurations in a system without gelation ($p=0.2$). (c) Time sequence of condensate configurations with gelation with $p=1$. Green: complex-rich phase; Black: gel phase; Transparent: complex-poor phase. (d) Average domain size cubed $\langle \bar{R}\rangle^3$ vs.~time displaying the expected behavior $\langle \bar{R}\rangle^3= \Omega t$. Inset: Coarsening prefactors $\Omega$ show strong dependence on the degree of gelation and decrease significantly when gel phase is present in condensates. Error bars show standard deviation of twenty repeated simulations.
}
\end{figure*}

\subsection{Brownian motion algorithm}

By assuming Stokes flow, Berry et al.~proposed an algorithm that couples the translational Brownian motion of condensates to hydrodynamic equations of motion corresponding to 3D Stokes flow \cite{Berry_2015}. When liquid droplets overlap, their cores are merged instantly with conservation of volume and center of mass (COM). While such an approach works well for simple liquid condensates, it encounters problems for gel-like domains, which behave rather differently when merging. 
In fact, recent evidence shows that the movements of gel-like condensates in a flow chamber may lead to granular structures, which subsequently grow into an interconnected network over time \cite{Ceballos_2022}. Furthermore, a full description of Brownian motion includes not only translational displacements, but also rotational ones, which may become significant when the merging aggregates are non-spherical, thus affecting the coarsening behavior and final morphology of the system. To investigate the non-equilibrium processes of both liquid and gel-like condensates under Brownian motion coalescence (BMC), we have developed an algorithm to effectively couple Brownian motion to the existing phase-field modeling framework as will be detailed next.

In our algorithm, once phase separation occurs, each complex-rich condensate, regardless of its geometry, is identified as an ``island'' using the Depth First Search (DFS) algorithm~\cite{Tarjan_1972}. The search algorithm assigns each identified island a label number, and the coordinates of every point within the island are stored for later computation. The COM of the island, $\vec r ^\mathrm{\ COM}_i$, in a domain with periodic boundary conditions, is subsequently determined. Assuming uniform mass distribution within each island, the COM is given by $\vec r ^\mathrm{\ COM} _i= \sum_{j=1}^{N_i} \vec r_{ij}/{N_i}$, where $\vec r_{ij}$ is the position of the $j$th voxel in island $i$, while $N_i(t)$ is the total number of voxels in island $i$.  [It is noteworthy that this method will fail for islands spanning across the periodic boundaries, an issue which can be circumvented by applying an algorithm from Bai and Breen in both 2D and 3D \cite{Bai_2008}.] Given the total (dimensionless) volume $N_i$ of the $i^{th}$ island, we define an effective radius of the island via $\bar{R}_i(t) \equiv \left(3N_i/4\pi\right)^{1/3}$. Upon the identification of all islands, a local velocity field $\vec v_i$, which is non-zero only in regions spanned by the $i^{th}$ island and zero elsewhere, is assigned. The effective advection velocity is then constructed via $\vec v = \sum_{i=1}^N \vec v_i$, where $N(t)$ denotes the total number of islands at a given time.  The advection of any field $\Psi$, including $X$, $Y$, $Z$ and $\phi$, is then achieved by evolving the continuity equation:
$\partial \Psi(\vec{r}, t)/\partial t+\nabla \cdot \left(\Psi \vec{v}  \right)=0$.
To numerically solve this equation, we use an upwind scheme to achieve numerical stability \cite{Smolarkiewicz_1983,Smolarkiewicz_1984}. This method ensures the conservation of buffer solution to achieve relative errors smaller than $1\times 10^{-6}$ after more than $1\times 10^7$ time steps in all the simulations reported in this work. 

Now, Brownian motion of the condensates has contributions from both translational and rotational displacements. To construct the velocity vectors $\vec v_i$ in 3D, we first consider translational motion. The island $i$ is displaced in a random direction every $m^{th}$ time step by a distance $\delta r_i = \sqrt{6D_i^{\mathrm{t}}m\Delta t}$, where $D_i^{\mathrm{t}}$ denotes the translational diffusion coefficient \cite{Berg_1984} and $m$ denotes the number of time steps in the iterative scheme. 
This assumes that the hydrodynamic drag forces and torques acting on each domain are isotropic, which is an oversimplification under some circumstances (e.g., slender objects experience smaller drag parallel to the long axis as compared to motion perpendicular to the long axis)~\cite{Brenner1963-part1, Brenner1964-part2}. However, in a system where coarsening and merging of domains are present, the domains become less anisotropic over time, rendering the assumption physically justified.
Next, a random unit vector, $\hat{v}_i^\mathrm{t}$, is generated for each island, representing the displacement direction of the entire island. The uniform translational velocity field assigned to all points within the island is thus $\vec v_i^{\ t} = \frac{\delta r_i}{m\Delta t}\hat{v}_i^\mathrm{t}$. Assuming rigid body rotation, the angular displacement of each island every $m$ time steps is $\delta \theta _i = \sqrt{4D_i^{\mathrm{r}}m\Delta t}$, where $D_i^{\mathrm{r}}$ denotes the rotational diffusion coefficient \cite{Berg_1984}. A unit vector through the COM of the island, $\hat{v}_i^\mathrm{r}$, is randomly generated as the axis of rotation. At each point within the island, a unit tangential velocity vector is computed as $\hat v_{ij} ^r = \hat{v}_i^\mathrm{r} \times \vec r^{\ d}_j$, where $\vec r^{\ d}_j$ denotes the shortest vector from $\vec r_{ij}$ to the rotation axis. The magnitude of the tangential velocity in turn is proportional to the distance from the rotation axis, i.e., $\vec v^\mathrm{\ r}_{i} = \omega _i |\vec r^{\ d}_j| \hat v_{ij} ^r$, where $\omega _i = \frac{\delta \theta _i}{m\Delta t}$ denotes the angular velocity, yielding the local velocity field $\vec v_i = \vec v_i^{\ t} + \vec v^\mathrm{\ r}_{i}$.

For an isolated spherical droplet in 3D Stokes flow, the translational and rotational diffusion coefficients are given by the Stokes–Einstein relations \cite{Koenderink_2003, Maldonado_2017} as $D^{\mathrm{t}}_i=\frac{k_{\mathrm{B}} T}{6 \pi \eta \bar{R}_i}$ and $D^{\mathrm{r}}_i=\frac{k_{\mathrm{B}} T}{8 \pi \eta \bar{R}_i^{3}}$, where $\eta$ denotes the viscosity. This results in the well-known relation between the diffusion coefficients $D_i^{\mathrm{r}} = 3D_i^{\mathrm{t}}/(4\bar{R}_i^2)$. However, the physical properties of the condensate and its surrounding medium are often non-ideal, leading to changes in size dependencies of $D^\mathrm{t}$ and $D^\mathrm{r}$. The formulation above allows free modifications of diffusivities, making it suitable for investigations of the scaling behavior in various systems, which we will elaborate using 2D simulations in the next Section.

\section{Results}

\subsection{Coarsening behavior without Brownian motion}

First, systems without Brownian motion are studied.
In Fig.~\ref{fig:3D_noBM}a, following the approach detailed in our companion article~\cite{Zhang_part1}, the phase diagram of a representative system, in which the gelation region partially overlaps with the LLPS one, is shown.  
For illustrative purposes, the degrees of polymerization are set to $r_x=r_y=r_z=1$; modifying these values lead to shifts in the phase boundaries without affecting the main features of the coarsening process. As discussed in the companion article~\cite{Zhang_part1}, the partial overlap between the LLPS and gelation regions imply that a significant fraction of systems prepared within the coexistence region will display gelation.  
In Fig.~\ref{fig:3D_noBM}b, we show the time evolution of a 3D system initialized with the compositions at $X_0 = Y_0 = 0.118$, and gelation parameters $p = 0.2$ and $Z^* = 0.4$, which yield a pure liquid-like system.
At early stages, complexes first form via the fast chemical reaction, subsequently phase separating from the buffer to form complex-rich condensates (green) via spinodal decomposition. As time progresses, the small liquid condensates undergo diffusion-limited coarsening (DLC or Ostwald ripening) to form larger condensates. In comparison, when increasing the fraction of cross-links formed in the complex to $p=1$ (Fig.~\ref{fig:3D_noBM}c), gel phase (black) evolves within the condensates 
as it meets the gelation criterion at $t = 2500$. The presence of gel phase slows down the reaction-diffusion process of the complex, leading to significantly lower coarsening rates and less spherical morphologies for condensates at late times. The coarsening rates of the system with various values of $p$ are quantified in Fig.~\ref{fig:3D_noBM}d. For all cases, coarsening of condensates asymptotically follow $\langle\bar{R}\rangle^3 = \Omega t$, agreeing with the Lifshitz–Slyozov–Wagner (LSW) theory~\cite{Lifshitz_1961, Wagner_1961}, where $\langle\bar{R}\rangle$ denotes the average domain or droplet size. When plotting the effective coarsening prefactors $\Omega$ in the inset, the liquid-like condensates ($pZ<Z^\ast$, i.e., $p=0,\ 0.2,\ 0.4$) display the highest coarsening rates. When higher fractions of cross-links form inside the condensates ($pZ>Z^\ast$, i.e., $p=0.6,\ 0.8,\ 1$), the coarsening rates significantly decrease with increasing $p$. 

\begin{figure}[h]
\includegraphics[width=0.85\linewidth]{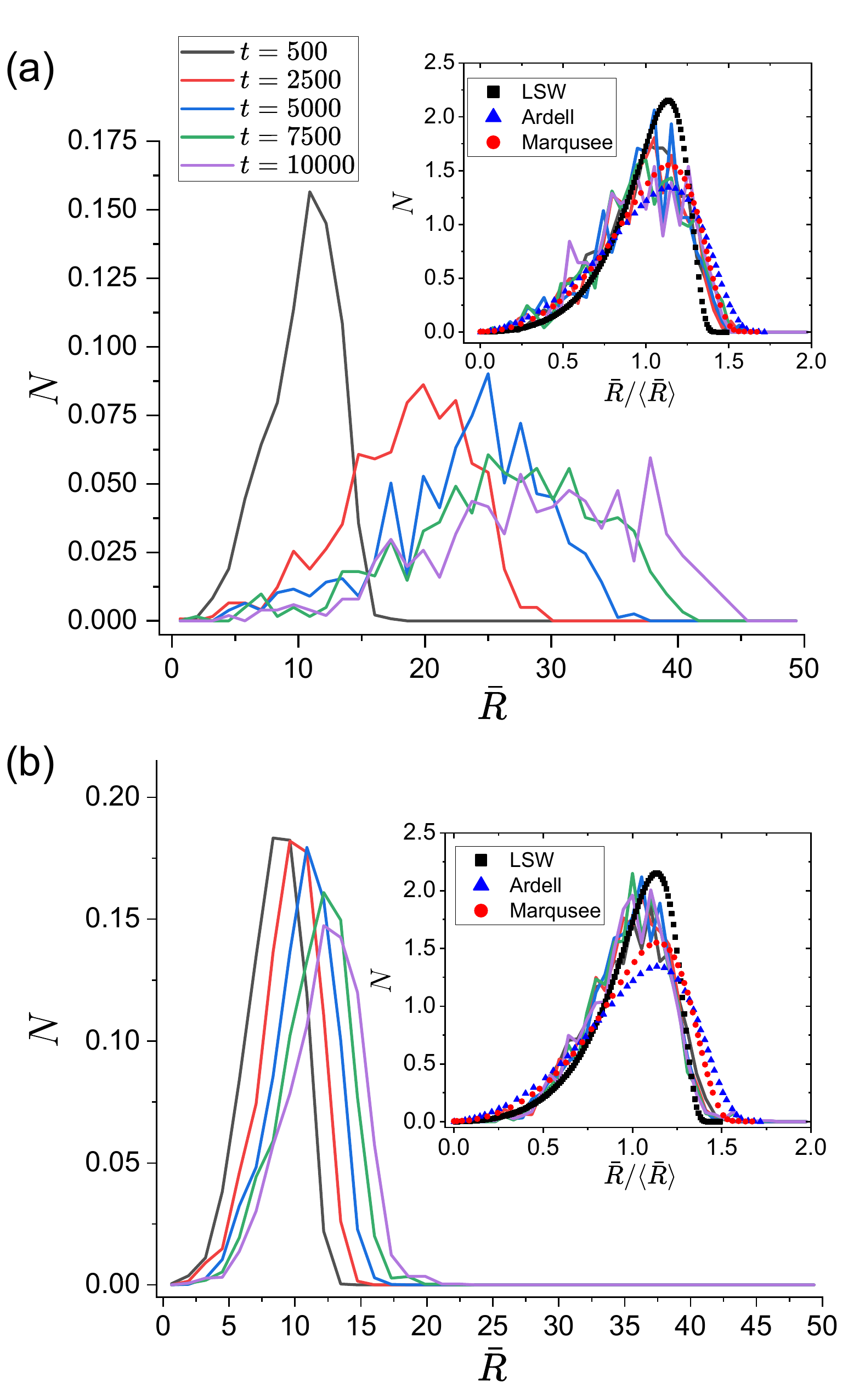}
\caption{\label{fig:size_DLC} Time-dependent size distributions in 2D systems without Brownian motion. (a) Liquid-like system. (b) Gel-like condensates. Insets: Scaled size distributions and comparison with analytic models from LSW \cite{Lifshitz_1961, Wagner_1961}, Marqusee \cite{Marqusee_1984} and Ardell \cite{Ardell_1990}. Liquid-like system shows good agreement with the Marqusee and Ardell models, while gel-like system behaves more like the LSW one. The distributions are obtained by averaging twenty random initial conditions for both liquid and gel-like systems.}
\end{figure}

We further investigate how gelation affects the size distribution of condensates. For a purely liquid system, LSW theory predicts the functional form of the droplet size distribution in a three-dimensional space in the limit of vanishing volume fraction of the condensate phase \cite{Lifshitz_1961, Wagner_1961, Baldan_2002, Dubrovskii_DLC_2011}. Marqusee et al.~and Ardell et al.~proposed models for two-dimensional size distribution with physically meaningful area fractions that are finite \cite{Marqusee_1984, Ardell_1990, Toral_1992}. We performed 2D simulations to assess the applicability of these condensate size-distribution models. For a finite area fraction of $0.21$, the simulation results of a liquid system have better agreement with Marqusee and Ardell's models than LSW theory (Fig.~\ref{fig:size_DLC}a). However, for a gel-like system (Fig.~\ref{fig:size_DLC}b), the coarsening of condensates is hindered by the physical cross-links, leading to a relatively stagnant change in the size distribution throughout the simulations. This resulting scaled size distribution (inset) is still in reasonable agreement with the LSW model.

\subsection{Coarsening behavior with Brownian motion in 2D}

\begin{figure*}[!ht]
\includegraphics[width=1\linewidth]{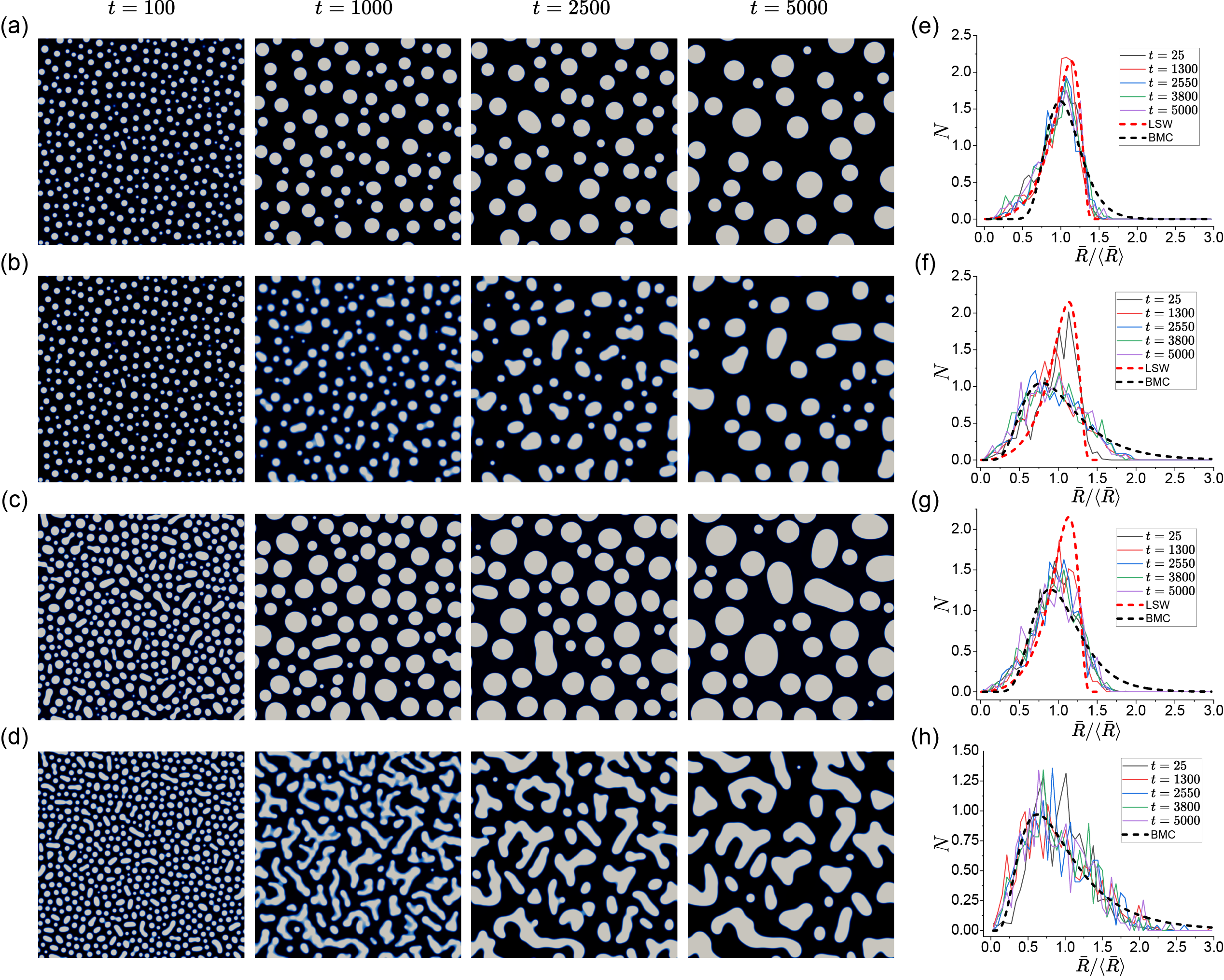}
\caption{\label{fig:BM2D_case1} Coarsening behavior of 2D systems with domain diffusivities given by the 3D Stokes-Einstein expressions $D^t = 0.25 /\bar{R}$ and $D^r = 0.1875/\bar{R}^3$. Time sequences of condensates: (a) Dilute (area fraction $=0.236$) and liquid-like; (b) Dilute and gel-like; (c) Dense (area fraction $=0.386$) and liquid-like; (d) Dense and gel-like. (e-h) Scaled size distributions of (a-d), respectively. The morphology and the size distribution of a system undergoing Brownian motion are strongly affected by gelation and the density.}
\end{figure*}

While the size dependencies of diffusion coefficients of spherical particles are well defined in 3D, there is no equivalent definition of Stokes-Einstein relation for 2D disks due to the ill-defined drag force and so-called Stokes' paradox~\cite{Saffman_1975, Berg_1984}. For practical measurements in finite quasi-2D systems, the standard Stokes-Einstein relation is only valid under certain conditions, and more often it breaks down~\cite{Lin_2DBM_1995, Liu_2DBM_2006, Sengupta2013}. To gain a better understanding of the effects of the size dependencies of diffusion coefficients on coarsening, we first conduct a systematic investigation of both liquid and gel-like condensates undergoing Brownian motion using 2D simulations. The translational displacement in 2D is $\delta r_i = \sqrt{4D_i^{\mathrm{t}} m \Delta t}$, and the angle of rotation becomes $\delta \theta _i = \sqrt{2D_i^{\mathrm{r}} m \Delta t}$ ~\cite{Hunter_2011}. The rotation axis is set to be perpendicular to the $xy$-plane, such that the rotation of 2D structures is either clockwise or counter-clockwise for a given island. The effective radius of each domain is now defined as $\bar{R}_i \equiv \sqrt{N_i/\pi}$.

Assuming that the effects of rotational displacements can be neglected during the coalescence of liquid condensates due to their spherical morphology, we first investigate how the size dependence of translational Brownian motion affects the overall coarsening rate within a simple picture of coalescence events driven by binary collisions.
As discussed by Siggia \cite{Siggia1979}, let us consider $N(t)$ spherical liquid droplets with radius $\bar{R}(t)$ per unit volume undergoing Brownian motion. The droplet volume fraction is defined as $f_V = \frac{4}{3}N(t)\pi \bar{R}^3(t)$. The rate of change of $N(t)$ due to binary droplet collision events can be expressed as $dN(t)/dt = -16\pi D^\mathrm{t} \bar{R}(t)N^2(t)$, or, written in terms of $\bar{R}(t)$, $\bar{R}(t)d\bar{R}(t)/dt= 4f_V D^\mathrm{t}$. If the translational diffusion follows $D^\mathrm{t} \sim \bar{R}^{-\alpha}$, we readily obtain $\bar{R}^{\alpha+2}(t) \sim f_V t$. Similar analysis in 2D leads to a similar scaling form $\bar{R}^{\alpha+2}(t) \sim f_A t$ (apart from logarithmic corrections), where $f_A$ denotes the area fraction. 
While these arguments provide guidelines for studying the coarsening rate of liquid condensates undergoing translational diffusion, the effects of rotational diffusion have been neglected based on the assumption of circular (spherical) shapes of the 2D (3D) condensates. However, as we will show, the gelation of the condensates leads to non-spherical morphologies, making such assumptions invalid. As a result, rotational diffusion becomes significant. Next, we conduct parametric studies on 2D systems to quantify how size-dependent translational and rotational diffusion coefficients affect the overall coarsening behavior. The data presented below were averaged over 20 independent simulations.

\subsubsection{Stokes-Einstein relation: $D^\mathrm{t} \sim 1/\bar{R},\ D^\mathrm{r} \sim 1/\bar{R}^3$}

\begin{figure*}[!ht]
\includegraphics[width=1\linewidth]{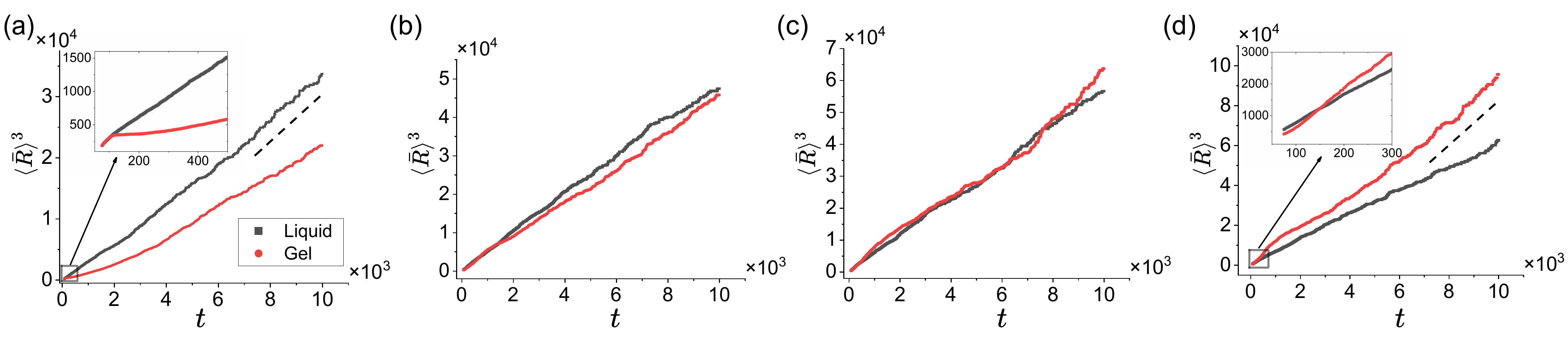}
\caption{\label{fig:BM2D_SEcross} Average domain size cubed $\langle \bar{R} \rangle^3$ vs.~time for 2D systems with $D^t = 0.25 /\bar{R}$ and $D^r = 0.1875/\bar{R}^3$ at various area fractions: (a) $0.236$; (b) $0.343$; (c) $0.365$; (d) $0.386$. The coarsening rate of the gel condensates overtakes the liquid ones at sufficiently large area fractions.}
\end{figure*}

We first consider a 2D system in which the domains execute Brownian motion with diffusivities following the 3D Stokes-Einstein relation $D^\mathrm{t} \sim 1/\bar{R}$ and $D^\mathrm{r} \sim 1/\bar{R}^3$.  As for a physical example of such systems, it has been shown that cylinders embedded within planar lipid bilayer membranes surrounded by a viscous bulk solvent exhibit such behavior when the length of the cylinder exceeds a characteristic length scale $\ell_0$ set by the ratio of membrane and ﬂuid viscosities~\cite{Levine2004-PRE}. 

Now, in a dilute liquid system with an area fraction of $0.236$ (Fig.~\ref{fig:BM2D_case1}a), both Ostwald ripening and Brownian motion contribute to the coalescence of the condensates. We observe from the simulations that BMC is the dominant mechanism at early stage, as the small condensates have larger displacements from Brownian motion. At late stage, the displacements of condensates caused by Brownian motion are significantly smaller due to the size dependencies, thus very few collisions can be observed, and the coarsening becomes more diffusion-limited. As a result, the late-stage size distribution does not fit well with the analytic steady-state size distribution for BMC while it shows better agreement with the LSW theory, which is also left-skewed (Fig.~\ref{fig:BM2D_case1}e). On the contrary, in a system where diffusion and reaction become strongly restricted by gelation, Brownian motion emerges as the main coarsening mechanism for the gel-like condensates. Unlike the instantaneous merging between liquid condensates, the collisions between gel-like condensates partially preserve the original morphologies of the condensates while the diffuse liquid-like interfaces merge. Many ``dimers'' and ``trimers'' are formed by the merging of circular condensates at $t = 1000$ (Fig.~\ref{fig:BM2D_case1}b). This morphology resembles the experimental observations on phase separated bilayer membrane systems that can form a ``sponge'' phase \cite{Kurita2022}. After the initial stage, the coarsening of the dimers and trimers leads to non-circular domains. The scaled size distribution of the gel-like system in Fig.~\ref{fig:BM2D_case1}f deviates from LSW theory but agrees well with the analytic distribution for a system dominated by BMC, suggesting that the suppression of the reaction-diffusion process by gel phase results in a change in the dominant coarsening mechanism from DLC to BMC.

Next, we study a more dense system with with an area fraction of $0.386$. We observe that the liquid-like mixture displays more frequent coalescence events caused by Brownian motion (Fig.~\ref{fig:BM2D_case1}c) when compared to the dilute mixture throughout the simulations. The scaled size distribution also reflects this observation by way of fitting better to BMC than LSW theory (Fig.~\ref{fig:BM2D_case1}g). For the gel-like system (Fig.~\ref{fig:BM2D_case1}d), many collision events can be observed at early stages, leading to the formation of connected domains with irregular morphologies. At late stage, however, the degree of anisotropy decreases as the domains continue to coarsen. 
The time evolution also shows direct evidence that branched gel-like domains possess an enhanced ability to recruit surrounding domains by rotation, reflected in a scaled size distribution predicted by the BMC distribution (Fig.~\ref{fig:BM2D_case1}h). 

The observations above indicate that the coarsening mechanism and size distributions of condensates are affected not only by the degree of gelation but also the area fraction. By plotting the average condensate radius cubed $\bar{R}^3$ vs.~time for various area fractions extracted from extended simulations in Fig.~\ref{fig:BM2D_SEcross}, we find that $\bar{R}\propto t^{1/3}$ holds at late stages of coarsening for all four systems. For the most dilute system (Fig.~\ref{fig:BM2D_SEcross}a), both liquid-like and gel-like mixtures have the same coarsening behavior at the earliest stages (inset), as it takes some time for the gel phase to form inside the condensates. Once the gel phase has formed, it hinders the reaction-diffusion process, leading to a sudden drop in the coarsening rate, deviating from the liquid-like curve. However, the coarsening rate of gel-like condensates keeps increasing, such that it becomes comparable to the liquid-like system one at late stages. 

More interestingly, as the area fraction is increased to $0.343$ (Fig.~\ref{fig:BM2D_SEcross}b), the coarsening rate of gel-like condensates keeps up with the liquid-condensates at early stages but still falling behind at intermediate and late stages.
By further increasing the area fraction to $0.365$ (Fig.~\ref{fig:BM2D_SEcross}c), we observe that gel-like condensates coarsen with the same rate as liquid-like condensates up to $t=9000$, and coarsen faster thereafter. Finally, in the simulations performed at the largest area fraction ($0.386$; cf.~Fig.~\ref{fig:BM2D_SEcross}d), a cross-over behavior is observed already at very early stages (inset). Thus, once the gel phase forms, BMC overtakes DLC as the main coarsening mechanism in gel-like systems leading to faster coarsening kinetics as compared to the liquid-like system at same area fraction, in striking contrast with systems without Brownian motion (cf.~Fig.~\ref{fig:3D_noBM}d).   

\begin{figure*}[!ht]
\includegraphics[width=1\linewidth]{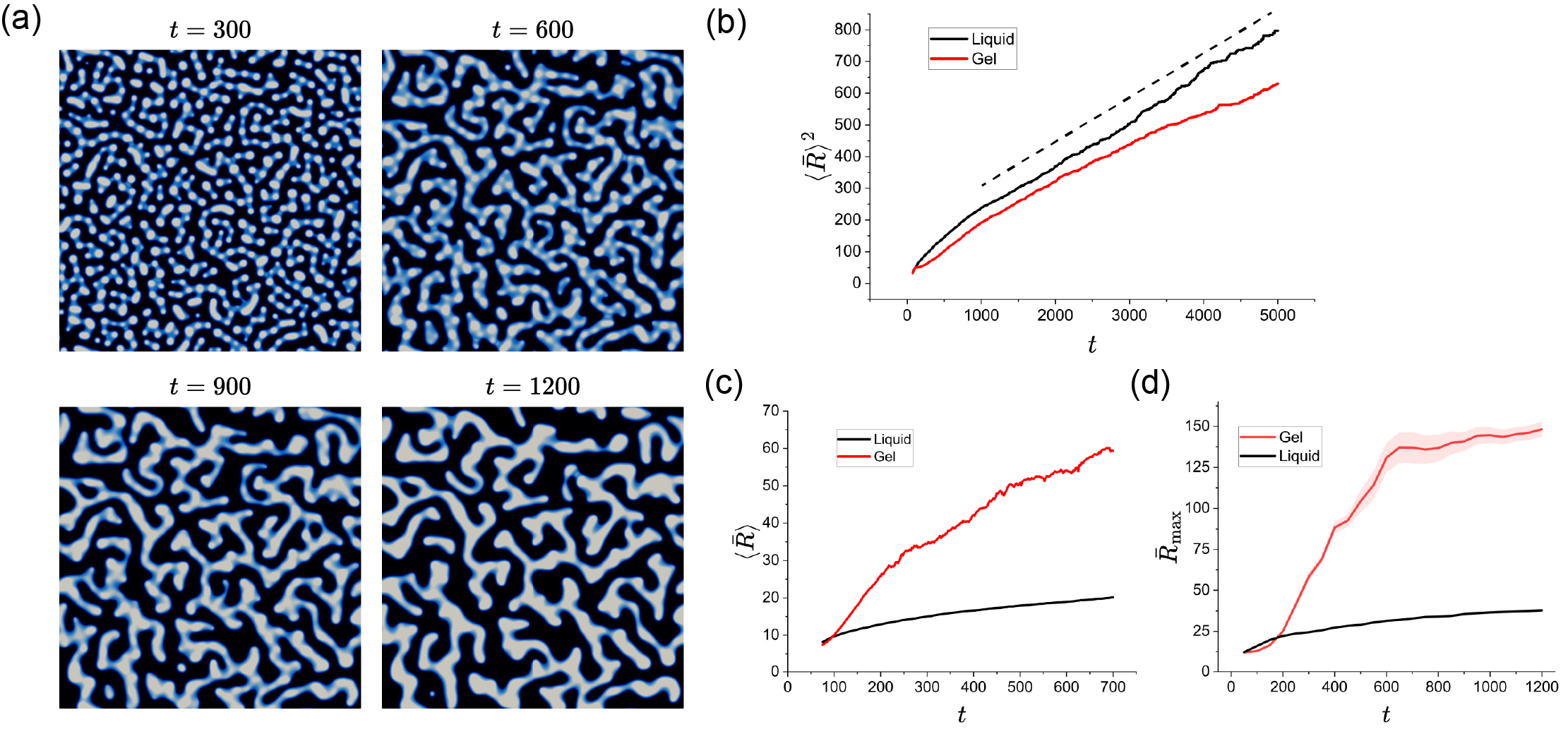}
\caption{\label{fig:BM2D_Saffman} Coarsening behavior of 2D system with Brownian motion with domain diffusivities given by the Saffmann-Delbr\"{u}ck \cite{Saffman_1975, Saffman_1976} expressions $D^t =0.025 \ln (L/\bar{R})$ and $D^r = 0.025/\bar{R}^2$. (a) Time sequences of gel-like condensates in a dense (area fraction of $0.386$) system. (b) Average domain size $\langle\bar{R}\rangle^2$ vs.~$t$ for the dilute (area fraction of $0.236$) system. The liquid-like system displays behavior $\langle\bar{R}\rangle^2 \sim t$ in agreement with the scaling analysis. (c) $\langle\bar{R}\rangle$ vs.~$t$ for the dense system. (d) Radius of the largest cluster $\bar{R}_{\text{max}}(t)$ for the two systems vs.~$t$. It is noteworthy that $\bar{R}_{\text{max}}(t)$ for the dense gel-like system rapidly outgrows its liquid counterpart before saturating due to finite-size effects. }
\end{figure*}

\subsubsection{Saffman-Delbr\"uck relation: $D^t \sim \ln (\ell_0/\bar{R}),\ D^r \sim 1/\bar{R}^2\ $}

Next we consider the case of domains embedded within a planar lipid bilayer membrane with domain sizes small compared to $\ell_0$~~\cite{Levine2004-PRE} with $\ell_0$ set equal to the simulation box length $L$. For a cylindrical domain of radius $\bar{R}$, Saffman and Delbr\"{u}ck derived the expressions $D^t \sim \ln (\ell_0/\bar{R})$ (corresponding to an effective exponent $\alpha \simeq 0$) and $\ D^r \sim 1/\bar{R}^2\ $~\cite{Saffman_1975, Saffman_1976}.  The predicted scaling behavior for $D^t $ was later verified by measuring the lateral diffusion of membrane proteins \cite{Ramadurai2009}. It was also utilized in a recent study which couples active flow fields to phase separated lipid membranes \cite{Arnold2023}. 

To illustrate how gelation affects the coarsening behavior in this system, we first note that while the liquid-like condensates continue to display circular domains, the gel-like ones with area fraction $0.386$ rapidly form interconnected structures as shown in Fig.~\ref{fig:BM2D_Saffman}a. The coarsening behavior of this system is also affected by the condensate area fraction. Figure \ref{fig:BM2D_Saffman}b shows that for a dilute system, liquid-like condensates coarsen faster than gel-like condensates and display behavior consistent with the predicted scaling behavior $\langle \bar{R} \rangle ^{2+\alpha}  \simeq \langle \bar{R} \rangle ^2 \propto t$ at late times. For the dense gel-like system, however, the formation of interconnected domains correlates with the rapidly increasing coarsening rate for $\langle\bar{R}\rangle$ displayed in Fig.~\ref{fig:BM2D_Saffman}c. More intriguingly, the radius of the largest cluster $\bar{R}_{\text{max}}(t) \equiv \max\{\bar{R}_i(t)\}$ for the dense gel-like system (cf.~Fig.~\ref{fig:BM2D_Saffman}d) outgrows its liquid counterpart already at the early stages of the coarsening process. We note that $\bar{R}_{\text{max}}(t)$ begins to plateau for $t>600$, beyond which $\langle \bar{R} \rangle$ is affected by finite-size effects. To better understand the respective roles of translational and rotational diffusion on coarsening rates, we next consider a scenario in which only rotational diffusion is active.

\subsubsection{No translation: $D^\mathrm{t} = 0$}

\begin{figure*}[ht!]
\includegraphics[width=1\linewidth]{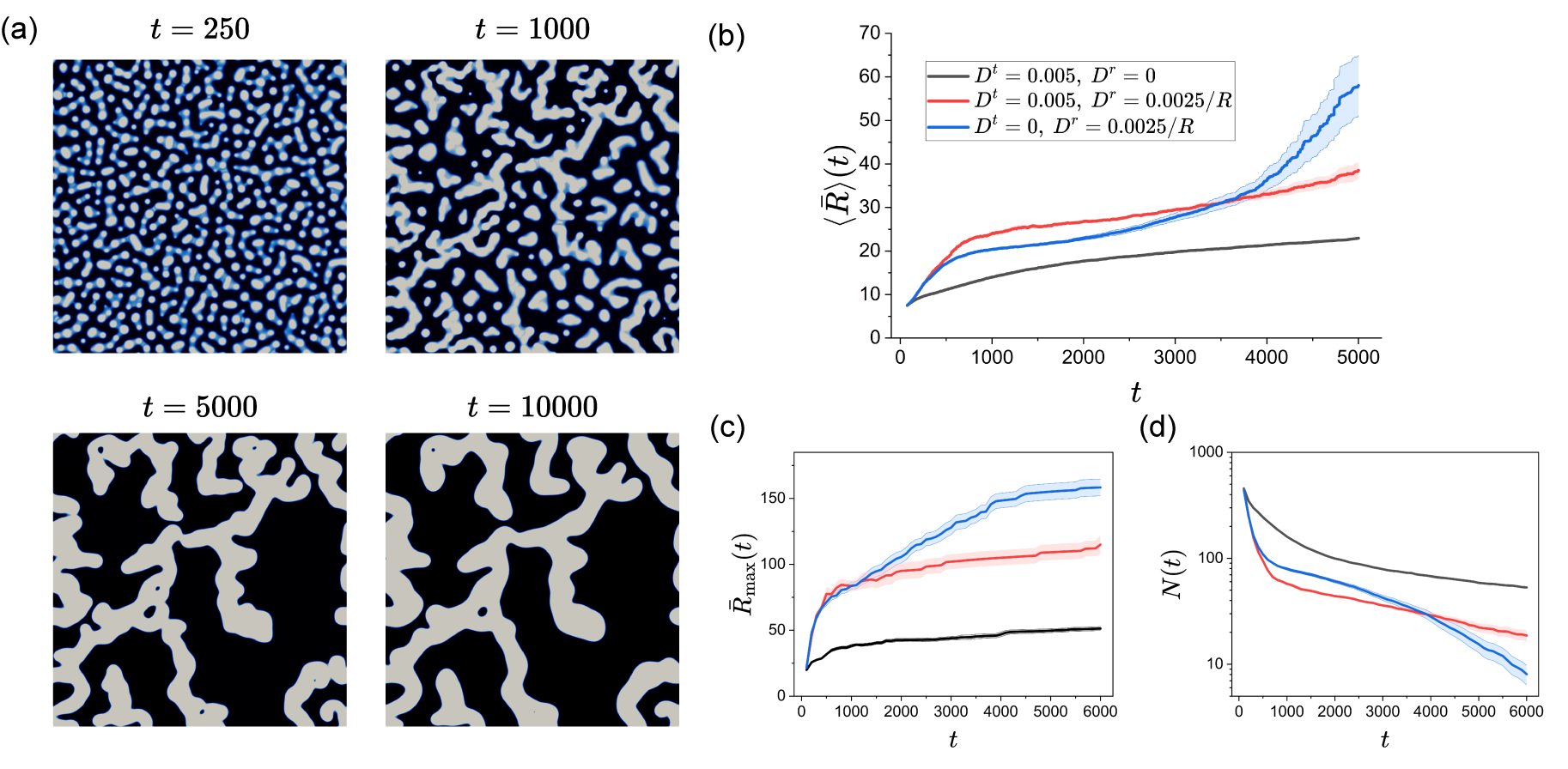}
\caption{\label{fig:BM2D_Dr_scaling}  (a) Snapshots from representative 2D simulations with Brownian motion, assuming $D^\mathrm{t}=0$ and $D^\mathrm{r} = 0.0025/\bar{R}$. [Note that $D^\mathrm{t}$ is set to zero at $t=250$ once the condensates have formed.] (b) Plot of $\langle \bar{R} \rangle$ vs.~time for three different combinations of size-dependent translational and rotational diffusivities. The coarsening rate for the case $(D^\mathrm{t} =0,D^\mathrm{r} = 0.0025/\bar{R})$ corresponding to rotational diffusion only (blue triangles) shows a rapidly increasing behavior consistent with the scaling analysis in the main text which predicts a finite-time singularity. (c) Radius of the largest cluster $\bar{R}_{\text{max}}(t)$ vs.~time. (d) Number of domains $N(t)$ vs.~time.}
\end{figure*}

To this end, we consider a dense gel system with $f_A = 0.386$ and set translational Brownian motion to zero at $t=250$ after the emergence of numerous condensates, while rotational Brownian motion remains active and non-zero through the simulations. The coarsening process is subsequently governed by the rotation of the domains and merging of domains upon colliding.  By setting $D^\mathrm{r} = 0.0025/\bar{R}$, the system evolves to two large interconnected domains at $t=5000$ (Fig.~\ref{fig:BM2D_Dr_scaling}a). The plot of $\langle \bar{R} \rangle$ vs.~$t$ (Fig.~\ref{fig:BM2D_Dr_scaling}b; blue line) in turn shows a rapid increase at intermediate times; as comparisons, we plot two other cases where $D^\mathrm{t} >0$ at all times. Interestingly, while the system exhibiting only rotational Brownian motion coarsens more slowly initially than a system experiencing both translational and rotational Brownian motion (Fig.~\ref{fig:BM2D_Dr_scaling}b; red line), it overtakes the latter one eventually. This can be understood as follows. At early times, the largest clusters in the two systems with rotational Brownian motion, as quantified by the radius of the largest cluster $\bar{R}_{\text{max}}(t)$ shown in Fig.~\ref{fig:BM2D_Dr_scaling}c, are comparable in size, while the number of clusters $N(t)$ (Fig.~\ref{fig:BM2D_Dr_scaling}d) decreases more rapidly in the case with both translational and rotational Brownian motion; this results in a larger average cluster size for the latter system.  At late times, however, $\bar{R}_{\text{max}}(t)$ in the system with only rotational Brownian motion surpasses that of the other case, while the ``surplus'' in $N(t)$ swiftly decreases, leading to the observed rapid increase in $\langle \bar{R} \rangle$ at large $t$. We again note that $\bar{R}_{\text{max}}(t)$ begins to plateau for $t>4500$, beyond which $\langle \bar{R} \rangle$ is dominated by finite-size effects. To better understand the emergence of elongated domains with growing aspect ratios in the simulations with $D^\mathrm{t}=0$ and the resulting large domain coarsening rates, we next perform a scaling analysis of a simplified system.

\subsection{Scaling analysis of coarsening driven by purely rotational Brownian motion}

Consider a 2D system of $N(t)$ needle-shaped gel condensates of length $L(t)$ and constant width $b$, accounting for the largest clusters in the system, in a square domain of total area $A$. The condensates occupying an area fraction $f_A=N(t) L(t) b/A$ are free to rotate but not translate.
Now, the average separation between two nearest neighbor condensates $\bar{L}$ is given by $ \bar{L}(t) \sim \sqrt{A/N(t)}$ or $ \bar{L}(t) \sim \sqrt{L(t) b/f_A}$. The typical time $\tau$ it takes for two such condensates to rotate and collide can be estimated from $(\Delta \theta)^2 \sim (\bar{L}/ L)^2 \sim D^\mathrm{r} \tau$, which leads to $\tau \sim \bar{L}^2/(L^2 D^\mathrm{r}) \sim b/(f_A L D^\mathrm{r})$. [Note that collisions will occur in this model only when $ L(t) \gtrsim \bar{L}$.]
Upon merging of the two condensates, the change in the area is then $dA_{\rm{cond}}/dt = b d L/dt \approx \Delta A_{\rm{cond}}/\tau \approx L(t) b/\tau \sim f_A L^2(t) D^\mathrm{r}$, or $bd L (t)/dt \sim f_A L^2(t) D^\mathrm{r}$. Since $\bar{R}_{\text{max}}(t) \sim \sqrt{L(t)b}$, we have $b^2d \bar{R}_{\text{max}} (t)/dt \sim f_A \bar{R}_{\text{max}}^3(t) D^\mathrm{r}$. Hence, with $D^\mathrm{r} =D_0 \bar{R}_{\text{max}}^{-\beta}$, we readily obtain $\bar{R}_{\text{max}}^{\beta-3}(t) d \bar{R}_{\text{max}}/dt \sim D_0f_A/b^2$. We first note that when $\beta = 3$ (corresponding to the Stokes-Einstein model), $\bar{R}_{\text{max}}(t) \sim D_0f_A t/b^2$, suggesting that rotational Brownian motion may indeed accelerate coarsening of gel-like domains relative to liquid-like ones at large area fractions, in qualitative agreement with our observations in Fig.~\ref{fig:BM2D_SEcross}. Next, when $\beta = 2$ (corresponding to the 2D Saffman-Delbr\"{u}ck model), $\bar{R}_{\text{max}}(t) \sim \exp(D_0f_A t/b^2)$ indicative of rapid coarsening, consistent with the data for the dense gel system shown in Fig.~\ref{fig:BM2D_Saffman}d. We also note that for the case $\beta = 1$ the scaling analysis suggests that there is a finite-time singularity $\bar{R}_{\text{max}}(t) \sim \bar{R}_{\text{max}}(0)/\left(1-D_0 \bar{R}_{\text{max}}(0)f_At/b^2\right)$, with $\bar{R}_{\text{max}}(t)$ diverging as $t \rightarrow t^* = b^2 / (D_0 \bar{R}_{\text{max}}(0)f_A)$. Furthermore, when $t \ll t^*$, $\bar{R}_{\text{max}}(t) \sim \bar{R}_{\text{max}}(0) + D_0 \bar{R}_{\text{max}}^2(0) f_At/b^2$, consistent with the linear trend seen in Fig.~\ref{fig:BM2D_Dr_scaling}c at intermediate times. A similar scaling analysis of a 3D gel system with cylindrical condensates of length $L(t)$ and cross-sectional area $b^2$ yields $\bar{R}_{\text{max}}^{\beta -5} d \bar{R}_{\text{max}}/dt  \sim D_0 f_V ^{2/3}/b^4$. Intriguingly, for the 3D Stokes-Einstein model with $\beta=3$, the simple model again predicts a finite-time singularity with $\bar{R}_{\text{max}}(t) \sim \bar{R}_{\text{max}}(0)/(1-D_0 \bar{R}_{\text{max}}(0) f_V^{2/3}t/b^4)$.  

It is instructive to redo the above analysis for the case of elongated domains with a {\it{fixed}} aspect ratio $q=L(t)/b(t)$ such that $\bar{R}_{\text{max}}(t) \sim  L(t)/\sqrt{q}$. In this case, $ \bar{L}(t) \sim L(t)/\sqrt{qf_A}$, and the condition that rotational collisions occur $ L(t) \gtrsim \bar{L}$ becomes $ qf_A \gtrsim 1$. Now, assuming that $ qf_A \gtrsim 1$, repeating the steps above leads to the result $\bar{R}_{\text{max}}^{\beta}(t) \sim qf_A t$, while in 3D the corresponding expression becomes $\bar{R}_{\text{max}}^{\beta}(t) \sim (q^2f_V)^{2/3} t$ with $ q^2f_V \gtrsim 1$.  It is interesting to note that since $\beta=2$ ($\beta=3$) for the 2D Saffman-Delbr\"{u}ck (3D Stokes-Einstein) model, the corresponding coarsening exponent of 1/2 (1/3) emanating from the rotational diffusion of the fixed aspect ratio domains coincides with that obtained for purely translational diffusion.

\begin{figure*}[!ht]
\includegraphics[width=1\linewidth]{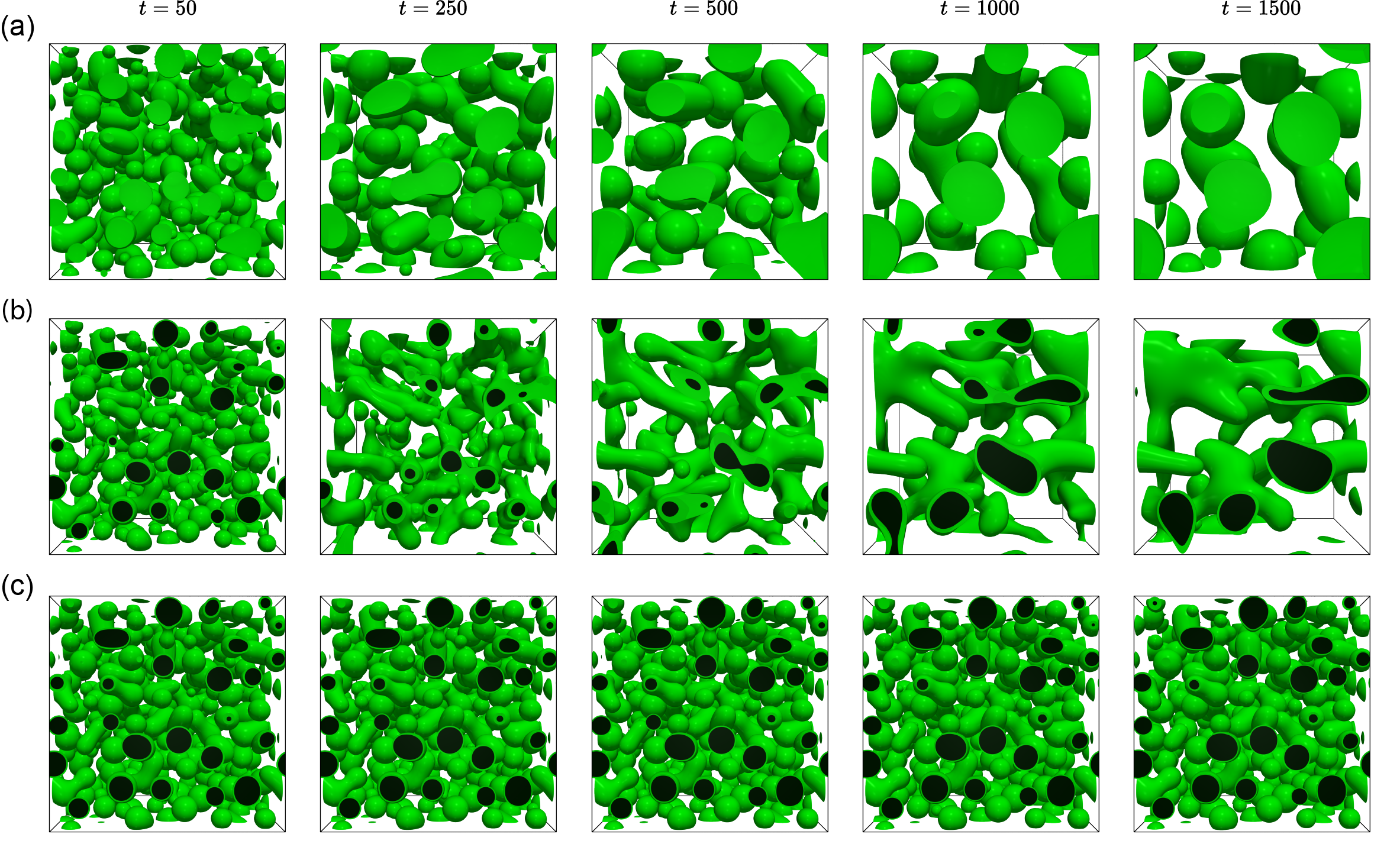}
\caption{\label{fig:BM3D} Snapshots of representative 3D simulations. (a) Liquid-like condensates with translational and rotational Brownian motion given by the Stokes-Einstein expressions $D^\mathrm{t}= 20/\bar{R}$ and $D^\mathrm{r} =15/\bar{R}^3$ form spherical domains. (b) Gel-like condensates with $D^\mathrm{t}= 20/\bar{R}$ and $D^\mathrm{r} =15/\bar{R}^3$ rapidly form a single connected domain. (c) Gel-like condensates without Brownian motion show very little coarsening over the entire time sequence. }
\end{figure*}

\subsection{Coarsening behavior with Brownian motion in 3D}

The algorithm can be readily extended to 3D systems. To compare how gelation and Brownian motion affect the morphology, we employ the same initial condition to study three cases, namely liquid-like and gel-like condensates with translational and rotational Brownian motion given by the Stokes-Einstein expressions $D^\mathrm{t}= 20/\bar{R}$ and $D^\mathrm{r} =15/\bar{R}^3$, and gel-like condensates without Brownian motion. First, in the liquid-like system (Fig.~\ref{fig:BM3D}a), small condensates undergo collisions to form larger condensates, which subsequently coarsen over time via BMC. The gel-like system, in contrast, forms irregular granular structures when small condensates are brought into contact by Brownian motion (Fig.~\ref{fig:BM3D}b). The elongated condensates keep evolving until they form a single interconnected domain ($t \geq 1000$). Experiments on gel-like systems report similar interconnected domains preserving the shape of individual spherical condensates after they collide, resembling the morphology we observed at early stages of gel-like condensates undergoing BMC \cite{Ceballos_2022}. Finally, when the same gel-like system evolves without Brownian motion (Fig.~\ref{fig:BM3D}c), instead of forming a gel network, the condensates become arrested in their original positions and coarsen extremely slowly via Ostwald ripening. 

\begin{figure}[ht]
\includegraphics[width=0.9\linewidth]{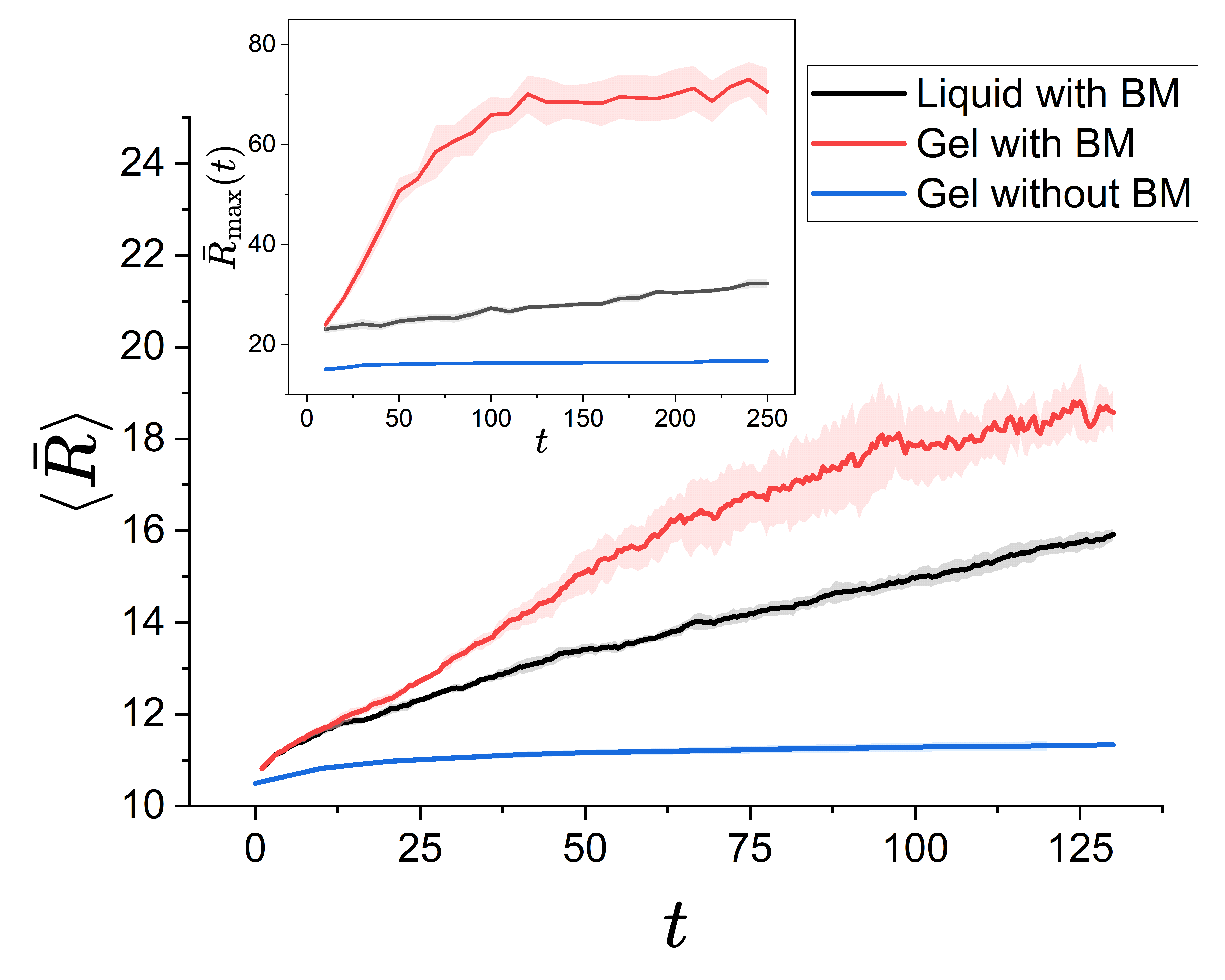}
\caption{\label{fig:scaling_3D} Average domain size $\langle \bar{R} \rangle$ vs.~time for the 3D simulations shown in Fig.~\ref{fig:BM3D}. Gel-like condensates with Brownian motion coarsen the fastest. Inset: Radius of the largest cluster $\bar{R}_\mathrm{max}$ vs.~time. For gel-like condensates with Brownian motion, the largest cluster spanning the simulation box appears around $t \approx 125$, beyond which $\langle \bar{R} \rangle$ is dominated by finite-size effects.}
\end{figure}

More quantitatively, next we compare the coarsening rates of the three systems by plotting $\langle \bar{R} \rangle$ vs.~time in Fig.~\ref{fig:scaling_3D}, averaged over five independent simulations each. It can be seen that gel-like condensates with Brownian motion coarsen the fastest. Perhaps more strikingly, as demonstrated in the inset, the largest cluster in the gel-like system with Brownian motion displays growth kinetics which greatly exceeds those of the liquid-like system with Brownian motion; in fact, the largest cluster spanning the simulation box appears around $t \approx 125$, beyond which $\langle \bar{R} \rangle$ is dominated by finite-size effects.  It should be stressed that we have chosen not to perform further quantitative analysis using 3D simulations due to the computational cost to achieve better statistical significance. The simulations reported herein are conducted using graphic processing units (GPUs), which are highly efficient in performing grid-based calculations. However, the Brownian motion algorithm, specifically the identification of the condensates using DFS, cannot be truly parallelized on GPUs. The serial tasks of DFS and velocity field calculations are thus being assigned to central processing units (CPUs) for every Brownian motion calculation. This in turn causes significant slowdown due to the time-consuming memory copying steps in large 3D systems. It is a part of our future studies to efficiently simulate systems with more than four components and one chemical reaction in 3D.

\section{Conclusions}
In this work, we have formulated a thermodynamically consistent phase-field modeling framework that combines the previously derived thermodynamic model with kinetic equations to study spatio-temporal evolution of macromolecular mixtures that can undergo chemical reactions, LLPS and gelation concurrently. We have observed significant morphological differences between the liquid and gel-like systems and characterized the coarsening behaviors by studying their size distributions and coarsening rates over time. When the system is liquid-like, the late-stage size distributions are consistent with the predictions from classic coarsening models \cite{Lifshitz_1961, Wagner_1961,Marqusee_1984,Ardell_1990} as expected. On the other hand, the gelation process within the condensates strongly hinders the chemical reaction and diffusion processes, thus significantly lowering the overall coarsening rate.

Furthermore, we have implemented a Brownian motion algorithm under the same computational framework to investigate how Brownian motion influences this kinetic process. By adding the translational and rotational displacements to each condensate, we have provided a means to effectively simulate a system in which coarsening is affected by both DLC and BMC. Using 2D simulations, we have demonstrated that different size dependencies of the Brownian motion of the domains -- in particular rotational -- may strongly affect the steady-state size distributions, coarsening rates and morphologies of the mixture. 
In addition, gelation of the condensates plays a key role not only in forming irregular and interconnected domains, but also accelerating the coarsening process in systems with high area fractions.

Through simulations and a simple scaling analysis, we demonstrated that the accelerated coarsening rates associated with gel-like condensates can be traced to the higher efficiency of elongated, high aspect ratio domains to experience further growth via collisions with nearby domains facilitated by rotational diffusion. In particular, the scaling analysis predicts that in some cases the domains may reach a macroscopic size in a finite time. This can be understood through percolation theory, which posits that the percolation threshold of a system of domains (both 2D and 3D) at fixed area/volume fraction is a strong function of the domain aspect ratio~\cite{Mertens2012,Rintoul1997}. For example, while disks (spheres) percolate at a critical area (volume) fraction  $f_A^* \simeq 0.67$ ($f_V^* \simeq 0.29$) in 2D (3D), spherocylinders (i.e., capsules) with aspect ratio $q=10$ percolate already at area fraction $f_A^* \approx 0.4$ ($f_V^* \approx 0.14$)~~\cite{Torquato2013}. Thus, if the coarsening process leads to domains with increasing aspect ratios, percolation will occur above a characteristic aspect ratio in both 2D and 3D, which will be reflected in the divergence of the average domain size. The resulting percolating gel structures resemble the bicontinuous interfacially jammed emulsion gels, or Bijels, which are formed by jamming of colloidal particles at the interface between two partially miscible fluids undergoing spinodal decomposition~\cite{Stratford_Bijel_2005}. Our simulations suggest an alternative pathway for stabilizing such non-equilibrium structures without using colloidal particles.

Finally, as we explored the passive Brownian motion in a system with undriven chemical reactions here, it would also be intriguing to consider active Brownian motion in a system with driven chemical reactions. It has been shown that liquid droplets containing bromine undergo self-propelling motion in an surfactant-rich oil phase~\cite{Thutupalli2011}. The bromination of the surfactant can induce Marangoni stresses, which then drives droplet motion. More recently, Testa et al.~show that chemically active liquid protein condensates can generate an activity-induced flow, driving the condensates to move and coalesce in a solution with pH gradient~\cite{Testa2021}. 
One may also couple the current formulation with the hydrodynamic equations or Model H~\cite{Hohenberg_1977} to capture more detailed hydrodynamic effects. 
Our work sets up a firm theoretical and computational framework for studies into these problems and further, which we intend to do in the near future.

\begin{acknowledgments}
RZ and MPH were supported by the National Science Foundation (NSF) Materials Research Science and Engineering Center Program through the Princeton Center for Complex Materials (PCCM) (DMR-2011750). SM acknowledges the partial support of the National Science Foundation of China (NSFC) under the grant number 12272005.
\end{acknowledgments}

\section*{Author Declarations}
\subsection*{Conflict of Interest}
The authors have no conflicts to disclose.

\subsection*{Author Contributions}
R.Z., S.M., and M.P.H. designed research, performed research, analyzed data and wrote the paper.

\section*{Data Availability Statement}
The data that support the findings of this study are available from the corresponding author upon reasonable request.

\section*{References}
\bibliography{manuscript}

\end{document}